\documentclass[preprint,authoryear,12pt]{elsarticle}

\usepackage{epsfig}

\usepackage{amssymb}

\usepackage[ps2pdf,%
a4paper=true,%
breaklinks=true,%
colorlinks=true,%
pdfauthor={First Author et al.},%
pdftitle={Template for manuscripts in Advances in Space Research}%
]{hyperref}

\journal{Advances in Space Research}

\begin{document}

\begin{frontmatter}



\title{Approaches to relativistic positioning around Earth and 
error estimations}                                   


\author[a]{Neus Puchades}
\author[a,b]{Diego S\'aez}
\address[a]{Departamento de Astronom\'\i a y Astrof\'\i sica, 
Universidad de Valencia, 46100-Burjassot, Valencia, 
Spain.}
\address[b]{Observatorio Astron\'omico, Universidad de Valencia,
46980 Paterna, Valencia, Spain}

\ead{diego.saez@uv.es}

\begin{abstract}
In the context of relativistic positioning, the coordinates of a given user
may be calculated by using suitable information broadcast by a 4-tuple 
of satellites. Our 4-tuples belong to the Galileo constellation.
Recently, we estimated the positioning errors due to  
uncertainties in the satellite world lines (U-errors).
A distribution of U-errors was obtained, at various times, 
in a set of points covering a large region surrounding Earth. 
Here, the positioning errors
associated to the simplifying assumption that photons move 
in Minkowski space-time (S-errors) are estimated and compared with the 
U-errors. Both errors have been calculated for the same 
points and times to make comparisons possible.
For a certain realistic modeling 
of the world line uncertainties,
the estimated S-errors have proved to be smaller than the U-errors, which
shows that the approach 
based on the assumption that the Earth's gravitational field produces negligible 
effects on photons
may be used in a large region surrounding Earth.
The applicability of this approach --which simplifies numerical
calculations-- to positioning problems, and the usefulness of our 
S-error maps, are pointed out. 
A better approach, based on the assumption that 
photons move in 
the Schwarzschild space-time governed by an idealized Earth, is also 
analyzed. More accurate descriptions of photon propagation 
involving non symmetric space-time structures are not necessary for 
ordinary positioning 
and spacecraft navigation around Earth.

\end{abstract}


\end{frontmatter}

\parindent=0.5 cm


\section{Introduction}
\label{intro}


This paper focuses on the estimation of positioning errors in some
relativistic positioning systems (RPS).
Realizations of RPS must be based on general relativity (GR);
more specifically, the following theoretical scheme is to be
implemented:     

(i) Space-time is governed by an energy distribution including Earth, Sun
and other sources. It is described by a metric, which must be written 
in terms of certain coordinates, $y^{\alpha} $, being appropriate 
to deal with positioning.

(ii) In the absence of non gravitational forces, satellites and photons move 
in the aforementioned space-time as test particles; namely,
following geodesics. In particular, satellites follow time-like geodesics which 
may be parametrized by the proper times. 

(iii) Once the metric and the world lines of satellites and photons are known, 
any user may calculate its $y^{\alpha} $ coordinates by using the proper times 
broadcast by four satellites, the so-called 
emission coordinates $\tau^{A}$. 

(iv) These emission coordinates
and the satellite world lines --parametrized by the proper times-- set the  
satellite coordinates at emission, which are also the initial photon coordinates;
namely, the coordinates at the starting point
of the photons broadcasting information. 

(v) Among the photon geodesics it is 
possible to find four intersecting ones (one per satellite), and
the coordinates $y^{\alpha} $
of the resulting intersection event define the user position.

Any approach to RPS based on these general ideas requires additional 
assumptions and approximations. 

In our opinion, the estimation of positioning errors is more 
systematic in the RPS context.
Instead of applying a correction for each effect 
leading to errors, let us systematically proceed taking into account 
the following points:

(1) We define {\em nominal satellite world lines}, which are 
appropriate time-like geodesics of the RPS space-time. These geodesics 
would be the true satellite world lines in the absence of perturbations

(2) In a given RPS, there are non gravitational perturbations 
such as solar winds, radiation pressure, and so on, as well as
gravitational perturbations due to the energy sources which have not been 
taken into account to fix the space-time structure. All these perturbations 
would produce growing deviations with respect 
to the nominal world lines

(3) These deviations will not be estimated but controlled; namely,
the satellite world lines will be corrected as soon as the amplitude 
of their deviations reaches a certain limit value
 
(4) The amplitude evolution and the statistical character of the 
deviations will be determined from the analysis of many data, 
which may be obtained by 
measuring deviations over an extended period (many satellite periods)
 
(5) Once nominal world lines and statistical realizations of the deviations 
are available,
the nominal world lines will be used to calculate the user position,
and the deviations to estimate errors. These 
errors will be hereafter called U-errors since they are due to 
uncertainties in the satellite world lines

(6) The best RPS would be obtained taking into account all the sources 
contributing to the gravitational field. 
In such a case, only the non gravitational forces would produce 
deviations with 
respect to the nominal world lines and, consequently,
less corrections of the satellite motions would be necessary
to maintain the deviations smaller than the chosen limit amplitude.

(7) There are other positioning errors associated to 
the description of photon propagation (from the 
satellites to the user). These errors arise when some sources of the gravitational field are 
neglected and, consequently, the metric and the photon null geodesics
are not fully accurate.

Two approaches to relativistic positioning 
are considered in this paper, in both cases we look for the positioning 
coordinates and their errors.
These RPS are designed by assuming 
that the space-time has the Schwarzschild metric,
which corresponds to an ideal isolated static spherically symmetric Earth.
Schwarzschild space-time is asymptotically Minkowskian and, consequently, 
once the approach based on Schwarzschild metric is assumed,
one can say that, from a theoretical (physical) point of view, there are inertial (quasi inertial)
systems of reference. The origin of these references is located  
in the Earth's center and the spatial axes are arbitrary.
The simplest of these two approaches, hereafter called the 0-order RPS,
is based on the following assumptions:
(a) satellite world lines are
time-like geodesics of the Schwarzschild space-time (hereafter S-ST), and (b) photons follow
null geodesics in the Minkowski space-time (M-ST) asymptotic to the Schwarzschild
space-time. 
In a more accurate approach, hereafter called the 1-order RPS, both satellites and 
photons move in S-ST.  
Here, the accuracy of the 0-order RPS is quantitatively estimated, for the first time,
in an extended region surrounding Earth.
This estimation is based on the calculation of the S-errors, 
which are the differences between the positioning coordinates obtained 
in the 0 and 1-order RPS.

In \cite{puc14}, the U-errors were estimated inside a 
spherical region,  with radius $R = 10^{5} \ km$,
centered at point E. 
The spherical inertial coordinates of E were assumed to be 
$r_{E} = R_{\oplus}$,
$\theta_{E} = 60^{\circ}$, and $\phi_{E} = 30^{\circ}$,
where $R_{\oplus}$ is the Earth's radius. Hence, 
point E is on the Earth's surface. It is an arbitrary point and results do not depend on its
choice.
In this paper, other positioning errors (S-errors) are estimated inside the same sphere
to facilitate comparisons with the U-errors. This great region
around Earth is hereafter referred to as the E-sphere. 

We only consider the Galileo constellation,
whose satellites are being placed in orbit by the European Space Agency.
This GNSS (global navigation satellite system) has 27 satellites, which are uniformly distributed on three 
equally spaced orbital planes.
We have numbered the satellites in such a way that numbers 1 to 9, 10 to 18, 
and 19 to 27 correspond to consecutive orbital planes. Inside any of these planes,
satellites $n$ and $n+1$ occupy successive positions.         
The inclination of these planes is $\alpha_{in}=56$ degrees 
and the altitude of the circular 
orbits is $h=23222 \ km$; thus, the orbital period is about $14 \ h$.
See \cite{pas07} for details. Our nominal world lines are chosen to be 
Schwarzschild time-like geodesics with these circular orbits.

Let us make some comments about notation and units which will be 
taken into account in the whole paper.
Index $A$ labels the four satellites necessary for 
space-time positioning; any other Latin index runs
from $1$ to $3$, and Greek indexes from $1$ to $4$.
Quantities $G$, $M_{\oplus}$, $t$, and $\tau $ stand for the gravitation constant,
the Earth's mass, the coordinate time, and the proper time, respectively.
In our numerical codes, 
units are chosen in such a way that the speed of 
light is $c=1$; the kilometer is the unit of distance, and times are given 
in units of $10^{-5}/3 \ s$ (hereafter {\em code units}). In all the equations
we set $c=1$ and, finally, 
results (code outputs) are presented in appropriate arbitrary units of 
distance and time.

The paper is structured as follows:
the 0-order RPS is briefly described in Sect.~\ref{sec-2}. 
The 1-order RPS, is studied in 
Sect.~\ref{sec-3}. The S-errors
are analyzed in Sect.~\ref{sec-4}, where they are 
compared 
with previously estimated U-errors \citep{puc14}. 
Main conclusions are 
briefly summarized in Sect.~\ref{sec-5}, where a general 
discussion is also presented.

\section{Emission and inertial coordinates in the 0-order RPS}
\label{sec-2}   
 
In the 0-order RPS, positioning coordinates are inertial coordinates
in the M-ST asymptotic to S-ST and,
consequently, they will be called 
{\em inertial asymptotic coordinates or inertial coordinates $x^{\alpha}$}, as done in previous 
papers \citep{puc12,sae13,puc14}.

The inertial coordinates (user position) may be found
by using the satellite world lines and the 
emission proper times, excepting some cases in which the emission coordinates are
compatible with two user positions (bifurcation); in these cases,
a criterion --based on additional data-- is necessary to choose the true position
\citep{sch72,abe91,cha94,gra96,col11a,col12a,puc12}.
Bifurcation does not play a role in this paper.

In any asymptotic inertial reference,
calculation of the user position requires 
knowledge of the satellite world lines.
If the equations of these lines are expressed in terms of the proper time
parameter, the coordinates of their points are functions 
of the form $x^{\alpha}_{A} = x^{\alpha}_{A}(\tau^{A})$.
Whichever the satellite world lines may be, if photons move 
along null geodesics in M-ST
(from emission to user reception), the emission coordinates, $\tau^{A} $, 
and the inertial coordinates, $x^{\alpha}$, are related as follows:
\begin{equation}
\eta_{\alpha \beta} [x^{\alpha} - x^{\alpha}_{A}(\tau^{A}) ]  
[ x^{\beta} - x^{\beta}_{A}(\tau^{A}) ] = 0   \ , 
\label{inemis}
\end{equation}    
where $\eta_{\alpha \beta}$ is the M-ST metric tensor, whose non 
vanishing components are 
$\eta_{11}=\eta_{22}=\eta_{33}=1 $ and $\eta_{44} =-1 $.

Given a point $Q$ of the M-ST with coordinates $x^{\alpha}$,
Eqs.~(\ref{inemis}) may be numerically solved to get the 
emission proper times $\tau^{A}$, which would be the emission coordinates 
received by an user at $Q$.     
A multiple precision numerical code has been designed to 
perform this calculation --leading to $\tau^{A}$-- for an arbitrary 
$Q$. It is hereafter referred to as the 
XT-code. This code uses both
the satellite world line equations and the numerical Newton-Raphson 
method \citep{pre99}.

\cite{col10a} derived an analytical solution 
of Eqs.~(\ref{inemis}), which 
gives $x^{\alpha}$ in terms of $\tau^{A}$ for photons moving in M-ST.
A numerical code based on this solution --hereafter referred to as the TX-code--
has been designed and tested \citep{neu11,puc12,sae13,sae14,puc14}.
The analytical solution holds for arbitrary satellite world lines; 
hence, these lines may be identified with the nominal
world lines of four Galileo satellites (see Sect.~\ref{intro}),
whose equations may be written in  
terms of the asymptotic inertial coordinates and the proper time. These equations 
are given in Sect.~\ref{sec-3}.  

In practice, perturbations deviate
any satellite from its nominal world line and, consequently, 
there are U-errors, which were 
estimated --in \cite{sae13,puc14}-- under the 
assumption that the satellite world lines are systematically corrected 
to always have space (time) deviations --with respect to the 
nominal world lines-- which are smaller than $10^{-2} \ km$ ($10^{-2} $ code units of time)
[see \cite{puc14} for more details]. 

\section{Relativistic positioning in S-ST: the 1-order RPS}
\label{sec-3}   

Various concepts and techniques being useful to develop the 1-order RPS have been found in previous papers,
among them, we may point out the definition and uses of the world function 
\citep{syn31,bah01,bin08,mig07} and the time transfer function, 
the form of this last function in the 
S-ST \citep{tey08}, and a method to find the user position coordinates by using 
the time transfer function \citep{cad05,cad10,del11}.
Here, this last method is modified by using the analytical 
formula derived by \cite{col10a} --instead of numerical iterations-- to work 
with photons moving in M-ST

The Earth's center is at rest in the asymptotic M-ST; hence, the S-ST may be 
considered as a perturbation of the asymptotic M-ST with a static metric 
$g_{\alpha \beta} = \eta_{\alpha \beta} + s_{\alpha \beta}$,
where $\eta_{\alpha \beta}$
is the Minkowski metric, and $s_{\alpha \beta}$
are perturbation terms depending 
on $GM_{\oplus}/R$, where $R$ is the Schwarzschild radial coordinate. 

The S-errors are due to the lensing deviations produced by the 
1-order RPS gravitational field, and these deviations are integrated effects on
the full paths traveled by the photons (from satellites to users).
Since the length of these paths has an order of magnitude whose values are 
$\sim 10^{4} \ km $ for users on Earth and $\sim 10^{5} \ km $ for users 
on the surface of the E-sphere (see Sect.~\ref{intro}) and, moreover, 
all the admissible nominal
world lines must be very close among them (with deviations 
smaller than $10 \ m$), it is evident that the photon paths, 
the lensing integrated
effect, and the S-errors will be almost identical for any 
admissible choice of the 
nominal lines. 
On account of this fact, 
the nominal world lines of the Galileo satellites 
may be assumed to be time-like geodesics of the S-ST with 
circular orbits. This simple choice is good enough 
to compute S-errors and, 
moreover, it allows comparisons with the U-errors calculated 
in \cite{puc14}, where the same nominal world lines were 
used.

In the unperturbed M-ST,
the spatial location of a Galileo satellite $A$,
which moves along a circumference, requires 
three angles. Two of them, $\Theta $ and $\psi $, characterize 
one of the three orbital planes. These angles are 
constant. The third angle, $\alpha_{A} $, localizes the satellite
on its trajectory. 
It depends on time. All this was taken into account in \cite{puc12,puc14} 
to find  
the world line equations of the satellite $A$
to first order in the small dimensionless parameter
$GM_{\oplus}/R$, whose maximum value is $GM_{\oplus}/R_{\oplus} \simeq
6.94 \times 10^{-10} $. These equations are as follows:
\begin{eqnarray}
x^{1}_{A} &=& R \, [\cos \alpha_{A}(\tau) \cos \psi + \sin \alpha_{A}(\tau) 
\sin \psi \cos \Theta] \nonumber \\ 
x^{2}_{A} &=& -R \, [\cos \alpha_{A}(\tau) \sin \psi - 
\sin \alpha_{A}(\tau) \cos \psi \cos \Theta] \nonumber \\
x^{3}_{A} &=& -R \sin \alpha_{A}(\tau) \sin \Theta \nonumber \\
x^{4}_{A} &=&  \gamma \tau   \ ,
\label{satmot1}
\end{eqnarray} 
where the factor $\gamma $ and the angle $\alpha_{A}$ are given by the relations \citep{ash03,pas07}
\begin{equation}
\gamma = \frac {dt}{d\tau} = \Big( 1 - \frac {3GM_{\oplus}}{R} \Big)^{-1/2} \ 
\label{ttau} 
\end{equation}  
and 
\begin{equation}
\alpha_{A}(\tau) = \alpha_{A0} - \Omega \gamma \tau  \ ,
\label{satmot2} 
\end{equation}   
respectively. The last equation involves the 
satellite angular velocity
$\Omega = (GM_{\oplus}/R^{3})^{1/2} $, and the angle    
$\alpha_{A0}$ fixing the position of satellite $A$ at $\tau = x^{4} = 0$
(GNSS initial operation time). 
The chosen nominal world lines satisfy
Eqs.~(\ref{satmot1})--(\ref{satmot2}).

Let us now assume that a certain set of four emission coordinates, $\tau^{A}$, has been received.
By using these coordinates,
the user position may be found in both the 0 and 1-order RPS.
In the first case,
photons move in the asymptotic M-ST, 
whereas for the 1-order RPS,
photons move in the proper S-ST and calculations are performed up to first
order in $GM_{\oplus}/R$. Thus, two different positions are obtained,
which allow us to estimate the positioning S-error. In both cases,
Eqs.~(\ref{satmot1})--(\ref{satmot2})
allow us to compute the position of any Galileo satellite (coordinates 
$x_{A}^{\mu}$) for any proper time $\tau$. Hence, from the four 
emission coordinates $\tau^{A} $, we may calculate the positions of 
the four satellites, at emission times, in S-ST. These emission events 
define the initial common points of the two photon world lines (in M-ST and S-ST); for this reason,
these events are denoted $P_{IA}$, and $x^{\mu}_{IA}(\tau^{A})$ are their 
asymptotic inertial coordinates in S-ST.

We first assume that photons move in S-ST. Those emitted from the satellite 
$A$ initiate their motion at the point $P_{IA}$ whose coordinates 
$x^{\mu}_{IA}(\tau^{A})$ have been calculated from $\tau^{A}$. The user position is then the 
intersection of four null geodesics that pass through points $P_{IA}$ and
have appropriate propagation directions at these points. The coordinates of the 
intersection point $P_{S0}$ (user space-time position in S-ST) are denoted $x^{\mu}_{S0}$.
Since the propagation directions 
are unknown, a suitable method is necessary to find the intersection point; namely,
to localize the user. 
A first version of the method used here, which does not use the exact solution of
\cite{col10a}, was implemented by \cite{del11}. 
Let us now describe our version of this method,
which uses this exact solution. Both versions
are based on the fact that, 
for the geodesic passing through points $P_{IA}$ and 
$P_{S}$ (coordinates $x^{\mu}_{S}$), there is a relation between $  x^{4}_{S} - x^{4}_{IA} \equiv t_{S} - t_{IA}$
and the coordinates $x^{i}_{IA} \equiv \vec{R}_{IA}$ and 
$x^{i}_{S} \equiv \vec{R}_{S}$. This relation may be written 
in the form 
\begin{equation}
t_{S} - t_{IA} = T_{S}(\vec{R}_{IA}, \vec{R}_{S})  \ ,
\label{S1}
\end{equation}
where $t_{S} - t_{IA}$ is the time elapsed from emission to event $P_{S}$, and 
$T_{S}$ is the so-called time transfer function corresponding to S-ST. 

Let us now consider that the photons emitted from the points $P_{IA}$
--with coordinates $x^{\mu}_{IA}(\tau^{A})$-- follow null geodesics in the 
asymptotic M-ST. Four of these geodesics intersect at point 
$P_{M0}$ (user space-time position in M-ST), 
whose coordinates $x^{\mu}_{M0}$ are to be compared with $x^{\mu}_{S0}$.
As in the S-ST, the time elapsed to go --along a null geodesic--
from the emission event to 
a point $Q_{M}$ with coordinates ($x^{i}_{M}, t_{M}$)
may be written as follows:
\begin{equation}
t_{M} - t_{IA} = T_{M}(\vec{R}_{IA}, \vec{R}_{M})  \ ,
\label{S2}
\end{equation}   
where $T_{M}$ is the time transfer function
of the asymptotic M-ST. In this simple geometry, the null geodesics are 
straight lines and, consequently, 
\begin{equation}
T_{M}(\vec{R}_{IA}, \vec{R}_{M}) = |\vec{R}_{M} - \vec{R}_{IA}|  \ .
\label{S3}
\end{equation}   

Quantities ($\vec{R}_{M0}$, $t_{M0}$); namely, the user asymptotic inertial coordinates
$x^{\mu}_{M0}$ may be calculated, from the emission coordinates 
$\tau^{A} $, by using the TX-code (see Sect.~\ref{sec-2})
This code
calculates all the possible user positions corresponding to given 
$\tau^{A} $ coordinates. We can find either one or 
two positions (bifurcation). Hence,  
from the emission coordinates 
$\tau^{A}$, we have described suitable methods to calculate:
$x^{i}_{IA}$, $t_{IA}$, $x^{i}_{M0}$, and $t_{M0}$.
In the bifurcation case, there are two pairs 
($x^{i}_{M0}$, $t_{M0}$) which may be
separately considered. Once a pair has been fixed,
the positioning solution in M-ST (0-order RPS) is corrected to get the corresponding 
solution in the 1-order RPS; namely, in  
S-ST. This correction may be estimated as follows: first of all, for any point $Q_{M}$
of the null geodesic passing through the emission and reception events in 
M-ST, an associated point $P_{S}$ of the corresponding null geodesic in 
S-ST is defined by the relation $x^{\mu}_{S}(P_{S}) = x^{\mu}_{M} + \Delta x^{\mu}_{M} $.
Thus, Eq.~(\ref{S1}) may be rewritten as follows:
\begin{equation}
t_{M} + \Delta t_{M} - t_{IA} = T_{S}(\vec{R}_{IA}, \vec{R}_{M} + \Delta \vec{R}_{M}) \ .
\label{S4}
\end{equation}   
The right hand side of this equation may be expanded up to 
first order in $\Delta \vec{R}_{M}$ to get:
\begin{eqnarray}
&&t_{M} + \Delta t_{M} - t_{IA} = \, T_{S}(\vec{R}_{IA}, \vec{R}_{M}) + \nonumber \\ 
&&\frac {\partial T_{S}(\vec{R}_{IA}, \vec{R}_{S})} {\partial \vec{R}_{S}} 
\Big |_{\Delta \vec{R}_{M}=0}
\cdot \Delta \vec{R}_{M}    \ .
\label{S5}
\end{eqnarray} 
Since $\Delta t_{M}$ is a first order quantity, all the terms 
in Eq.~(\ref{S5}) must be developed up to this order;
hence, function $T_{S}(\vec{R}_{IA}, \vec{R}_{M})$ is to be expanded up to first order,
but the gradient involved in Eq.~(\ref{S5}) is only required at zero order,
since it is multiplied by the first order quantity $\Delta \vec{R}_{M}$.

At zero order, S-ST is identical to M-ST and, consequently, taking into account 
Eq.~(\ref{S3}) one finds  $T_{S}(\vec{R}_{IA}, \vec{R}_{M})
= |\vec{R}_{M} - \vec{R}_{IA}| + O(1)$; namely, the zero order approximation of 
the Schwarzschild time transfer function is $T^{(0)}_{S}(\vec{R}_{IA}, \vec{R}_{M})
= |\vec{R}_{M} - \vec{R}_{IA}| $. Hence, at zero order, the gradient of Eq.~(\ref{S5})
may be written in the form:
\begin{equation}
\frac {\partial T_{S}(\vec{R}_{IA}, \vec{R}_{S})} {\partial \vec{R}_{S}} 
\Big |^{(0)}_{\Delta \vec{R}_{M}=0} = \frac {\vec{R}_{M} - \vec{R}_{IA}} 
{|\vec{R}_{M} - \vec{R}_{IA}|} \ .
\label{S6}
\end{equation}

It has been proved that $T_{S}$  may be expanded as follows [see \cite{tey08} and references cited 
therein]:
\begin{equation}
T_{S} = |\vec{R}_{M} - \vec{R}_{IA}| + T^{(1)}_{S} + O(2) \ ,
\label{S7}
\end{equation}
with 
\begin{equation}
T^{(1)}_{S} = 2GM_{\oplus} \ln \Big[\frac {|\vec{r}_{M}|
+|\vec{r}_{IA}| + |\vec{r}_{M} - \vec{r}_{IA}|}
{|\vec{r}_{M}|
+|\vec{r}_{IA}| - |\vec{r}_{M} - \vec{r}_{IA}|} 
\Big]  \ ,               
\label{S8}
\end{equation}
where $r = |\vec{r}| = |\vec{R}| (1-GM_{\oplus}/|\vec{R}|) $ is the 
so-called radial isotropic coordinate [see \cite{mis73}]. 

From Eqs.~(\ref{S2}),~(\ref{S5}),~and~(\ref{S6})--(\ref{S8}), one easily gets
\begin{eqnarray}
&& \Delta t_{M} - \frac {\vec{R}_{M} - \vec{R}_{IA}} 
{|\vec{R}_{M} - \vec{R}_{IA}|} \cdot \Delta \vec{R}_{M} =  \nonumber \\
&& 2GM_{\oplus} \ln \Big[\frac {|\vec{r}_{M}|
+|\vec{r}_{IA}| + |\vec{r}_{M} - \vec{r}_{IA}|}
{|\vec{r}_{M}|
+|\vec{r}_{IA}| - |\vec{r}_{M} - \vec{r}_{IA}|} 
\Big]  \ . 
\label{S9}
\end{eqnarray} 
These equations must be particularized at point $\vec{R}_{M} = \vec{R}_{M0}$
and $t_{M}=t_{M0}$. After particularization,
there are four equations to find the four unknowns 
$\Delta \vec{R}_{M0}$ and $\Delta t_{M0}$ and, then, 
the positioning coordinates in S-ST are 
$\vec{R}_{S0} = \vec{R}_{M0} + \Delta \vec{R}_{M0} $ and 
$t_{S0} = t_{M0} + \Delta t_{M0}$. Each of the four Eqs.~(\ref{S9})
corresponds to a particular satellite (index $A$). The determinant of the
particularized system of four equations is
\begin{equation} 
D =            
\left| \begin{array} {cccc}                                                   
\frac{x^{1}_{I1} - x^{1}_{M0}} {|\vec{R}_{M0} - \vec{R}_{I1}|} & \frac{x^{2}_{I1} - x^{2}_{M0}} {|\vec{R}_{M0} - \vec{R}_{I1}|}
& \frac{x^{3}_{I1} - x^{3}_{M0}} {|\vec{R}_{M0} - \vec{R}_{I1}|} & 1 \\
                                   &                                    &                                    &   \\ 
\frac{x^{1}_{I2} - x^{1}_{M0}} {|\vec{R}_{M0} - \vec{R}_{I2}|} & \frac{x^{2}_{I2} - x^{2}_{M0}} {|\vec{R}_{M0} - \vec{R}_{I2}|}
& \frac{x^{3}_{I2} - x^{3}_{M0}} {|\vec{R}_{M0} - \vec{R}_{I2}|} & 1 \\  
                                   &                                    &                                    &   \\ 
\frac{x^{1}_{I3} - x^{1}_{M0}} {|\vec{R}_{M0} - \vec{R}_{I3}|} & \frac{x^{2}_{I3} - x^{2}_{M0}} {|\vec{R}_{M0} - \vec{R}_{I3}|}
& \frac{x^{3}_{I3} - x^{3}_{M0}} {|\vec{R}_{M0} - \vec{R}_{I3}|} & 1 \\  
                                   &                                    &                                    &   \\ 
\frac{x^{1}_{I4} - x^{1}_{M0}} {|\vec{R}_{M0} - \vec{R}_{I4}|} & \frac{x^{2}_{I4} - x^{2}_{M0}} {|\vec{R}_{M0} - \vec{R}_{I4}|}
& \frac{x^{3}_{I4} - x^{3}_{M0}} {|\vec{R}_{M0} - \vec{R}_{I4}|} & 1 \\  
\end{array} \right|
\ . 
\label{deter}
\end{equation}      
If this determinant vanishes for the emission coordinates $\tau^{A} $, 
there is no solution to the corresponding system of equations, which means that
positioning at the M-ST point $P_{M0}$ [with coordinates ($x^{i}_{M0},t_{M0}$)]
is not possible. Close to a point of this kind, the determinant must be small and,
consequently, the deviations $\Delta \vec{R}_{M0} $ and $\Delta t_{M0}$
are expected to be large.
Let us now raise the following question: are there emission 
coordinates leading to a vanishing determinant? 

Following \cite{puc14}, the value of $|D|$ is 
just $6V_{T}$, where $V_{T}$ stands for the volume of the {\em tetrahedron
formed by the tips of the four user-satellite unit vectors},
which have their common origin at point ($\vec{R}_{M0}, t_{M0}$)
in M-ST. See also \cite{lan99}, where this volume was related with 
the so-called dilution of precision (U-errors).
If the four unit vectors mentioned above correspond to 
generatrices of the same cone, their tips are on the same plane and,
consequently, the volume $V_{T} $ vanishes; hence, 
the answer to the above question is positive. In other words,
there are degenerate configurations leading to
vanishing $D$, in which, the user would see the four satellites on the same 
circumference.

If a spacecraft carries devices to get 
the line of sight of any visible Galileo 
satellite at emission time, the volume $V_{T}=|D|/6$ may be estimated
for any visible 4-tuple and,
consequently, 4-tuples leading to excessively small $|D|$ values may be rejected.
After improvements, this method for 4-tuple selection might help us 
to design a good autonomous spacecraft navigation system 
--up to distances around $10^{5} \ km$-- based, e.g., 
on Galileo satellites. 

The regions where the S-errors are expected to be too large must be located 
either close to 
points where $D$ vanishes, or at points located very far from the satellites.
At these last points, the four satellites 
are all in a small solid angle and the tetrahedron volume 
$V_{T}$ is expected to be small.
Our numerical estimates are in agreement with these expectations
(see next sections). The same is also valid for the U-errors estimated in 
\cite{puc14}.

\section{Numerical analysis}
\label{sec-4} 

In practice, the user --whose position is unknown-- receives four emission proper times
$\tau^{A} $, which may be used to get two sets of inertial coordinates (positions). 
One of them corresponds to null geodesics in M-ST and the other one to 
photon motions in S-ST. Nevertheless, in order to get a distribution of 
positioning S-errors, we may proceed as follows:
Given four satellites of the Galileo constellation, and a user position 
($x^{i}_{M0}$, $t_{M0}$) in M-ST, whose coordinates 
$x^{i}_{M0}$ correspond to a point of the E-sphere (see Sect.~\ref{intro}),
the XT-code (see Sect.~\ref{sec-2}) may be used to get the emission coordinates
$\tau^{A}$ which would be received by the chosen M-ST user. 
From these emission coordinates and the satellite world lines 
in S-ST [Eqs.~(\ref{satmot1})--(\ref{satmot2})], the initial photon positions 
$(x^{i}_{IA}, t_{IA})$ may be obtained and, then,                 
the S-errors at the M-ST chosen 
point ($x^{i}_{M0}$, $t_{M0}$) may be calculated by solving Eqs.~(\ref{S9})
for the unknowns $\Delta x^{i}_{M0}$ and $\Delta t_{M0}$. 
Quantities $\Delta_{R} = \Delta |\vec{R}_{M0}|$  and $\Delta_{t} = \Delta t_{M0}$ 
are good estimators of the 
S-errors.

\subsection{General considerations}
\label{sec-4-1} 

For a comparison with previous calculations of U-errors, 
which is necessary 
to decide if 
the 0 and 1-order RPS may be applied to a given positioning problem,
the determinant $D$ and the estimators $\Delta_{R} $ and $\Delta_{t} $ must be calculated
for appropriate satellite 4-tuples and users (space-time points).
These quantities have been found for a set of
$t=constant $ space-time hypersurfaces, in many points conveniently placed  
on four great spheres with different radii and concentric with Earth. These spheres are inside the E-sphere.
The same points and spheres were already considered to calculate U-errors (see \cite{puc14} 
for details).

We have considered   
many 4-tuples and hypersurfaces of constant time (hereafter given in hours),
but the main properties of the S-errors may be pointed out by using only the 4-tuple
2, 5, 20, and 23, and the hypersurface $t=19 $, as
done in the rest of this paper.

As done in \citep{puc14}, the
spheres are pixelized by using the HEALPIx package, which
was designed to build up 
temperature maps of the cosmic microwave 
background (CMB). Quantities $\Delta_{R} $ and $\Delta_{t} $ are
calculated for users located at the centers of the 
HEALPIx pixels and, then, pixels are colored (color bar) to show
the value taken by $\Delta_{R} $ or $\Delta_{t} $ in them.
Colors outside the bar may be used to mark special pixels;
in particular, pixels where the represented quantity is not defined or it satisfies some 
condition. 

Finally, 
the mollweide projection is used to show, in unique figure,
the whole pixelized colored sphere.
The frontal hemisphere is represented in the 
central part of the figure, and the opposite hemisphere is projected
on the lateral parts. The external edges of these parts
represent the same back semi-meridian.
HEALPIx-mollweide CMB maps may be seen, e.g., in 
\cite{ben13}.

Figs.~\ref{figu4} and~\ref{figu5} show HEALPIx-mollweide maps.
The interpretation of these Figures only requires:
the definition of the quantities represented in the maps, and the 
rules used to assign colors which are not in the bar (special pixels).
This information is given in the figure captions. Interpretations are 
detailed in the text.

In \cite{gor99}, the reader may find more details about
the HEALPIx pixelization, which is 
a hierarchical equal area isolatitude pixelization
of the sphere. 
The number of pixels is $12 \times N_{side}^{2}$, where 
the free parameter
$N_{side} $ takes on even natural values. 
All the HEALPIx pixels have not the same shape, they
are more elongated in the polar zones.
In our maps, the HEALPIx parameter $N_{side} $
is chosen to have $3072$ pixels and, then, 
the angular area subtended by each of these pixels is close 
to sixty four times the mean angular area of the full moon
($\sim 13.43 $ squared degrees). 
The angular separation between the directions associated to two 
neighboring equatorial pixels is 
$\Delta \alpha \simeq 0.1 \ rad$.
We have verified that this pixelization 
is adequate for our purposes.

For a given 4-tuple of satellites, positioning is only possible if the user 
may see the four satellites at the same time (visibility points). On account 
of this fact, we proceed as follows: (a) 
the users which do not see the four 
satellites (invisibility points) are identified, (b) 
the corresponding S-errors are not calculated and, (c) these users 
are properly marked in graphic representations.

\subsection{S-errors along typical radial directions}
\label{sec-4-2} 

\begin{figure*}[tb]
\begin{center}
\resizebox{.5\textwidth}{!}{%
\includegraphics{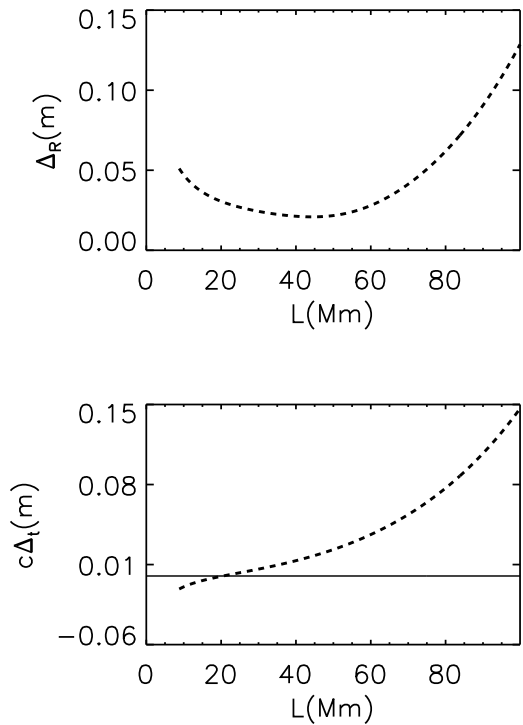}
}
\caption{The values of $\Delta_{R} $ (top) and $c\Delta_{t} $ (bottom) [in meters]
are represented
in terms 
of the distance $L$ to point $E$ in megameters. This representation corresponds 
to a typical direction without any $D=0$ point  
}
\label{figu1} 
\end{center}      
\end{figure*}

\begin{figure*}[tb]
\begin{center}
\resizebox{.5\textwidth}{!}{%
\includegraphics{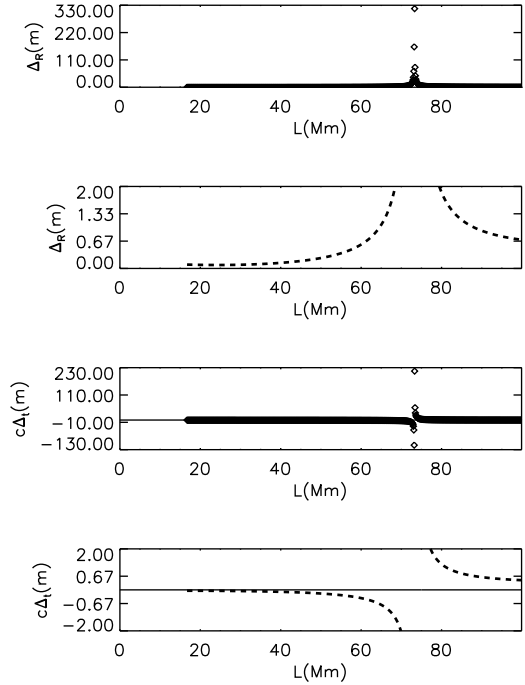}
}
\caption{In the top and middle-bottom panels,
the same representation as in Fig.~\ref{figu1}
is shown. Along the chosen direction quantity $D$ vanishes once.
The peak and the discontinuity are associated 
to the $D=0$ point.
Once the                  
$\Delta_{R} $ and $c|\Delta_{t}| $ values greater than two meters
have been eliminated,
the same representation 
as in Fig.~\ref{figu1} is repeated in the middle-top and bottom
panels. All the $\Delta_{R} $ and $c|\Delta_{t}| $ values 
smaller than two meters 
become so visible
}
\label{figu2} 
\end{center}      
\end{figure*}

\begin{figure*}[tb]
\begin{center}
\resizebox{.5\textwidth}{!}{%
\includegraphics{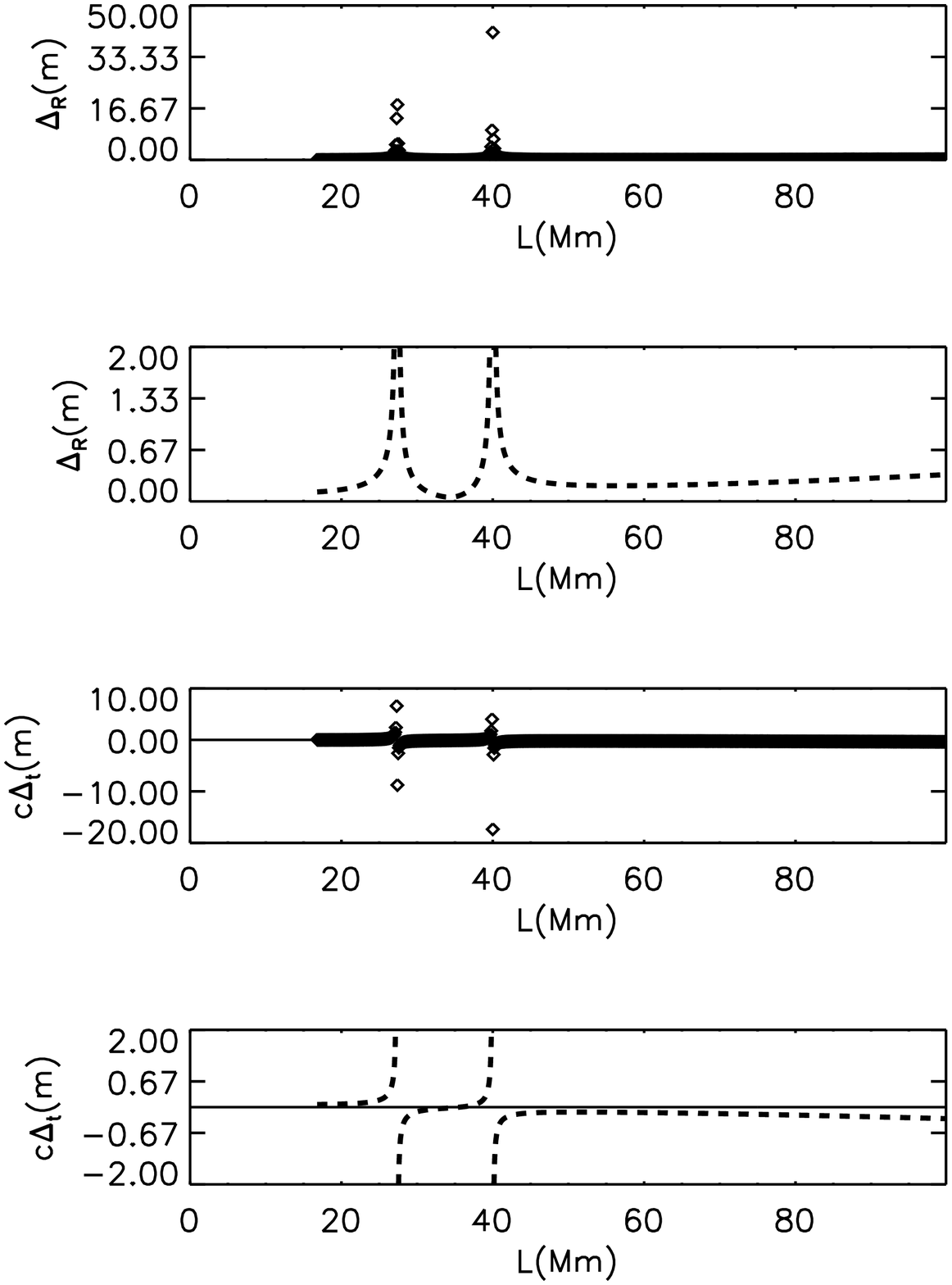}
}
\caption{Same representation as in Fig.~\ref{figu2}
for a typical direction where $D$ vanishes twice
}
\label{figu3} 
\end{center}      
\end{figure*}

Quantities $\Delta_{R}$ and $\Delta_{t}$ have been also calculated,
along many HEALPIx directions, at points whose distances $L$ to E range from $L=0 $ to $L=10^{5} \ km$.
The values corresponding to three characteristic directions are now presented. 
Results are displayed in Figs.~\ref{figu1}--\ref{figu3}.

The first direction (Fig.~\ref{figu1}) does not contain any $D=0 $ point and
we see that, for large enough $L$ values, the estimators $\Delta_{R}$ and $|\Delta_{t}|$
are continuous increasing functions
of $L$. From point E ($L=0$) to 
the starting point of the curve of Fig.~\ref{figu1}
($L \simeq 8800 \ km$), 
there is an invisibility 
segment since Earth --very close to the users in this segment-- 
hides one or more satellites of the chosen 4-tuple. In Figs.~\ref{figu2} and~\ref{figu3},
the length of the corresponding segment is $16700$ and $16800 \ km$,
respectively.

The second direction (Fig.~\ref{figu2}) contains only a $D=0 $ point
at $L \simeq 73300 \ km$.
In the top and the middle-bottom panels of this Figure, the position of the $D=0$ point is
the center of the $\Delta_{R}$ peak and the $\Delta_{t}$ discontinuity, respectively. 
Around the $D=0 $ point, quantities $\Delta_{R}$ and $|\Delta_{t}|$
are very large. In the middle-top and bottom panels, only
the values of these quantities smaller than $2 \ m$ have been represented; hence, in the region between the 
two dashed lines of these panels, $\Delta_{R}$ and $|\Delta_{t}|$ take on 
values greater than two meters. In this case, we easily see that values of 
two meters arise at distances (hereafter 2m-distances $\equiv$ $\Delta L_{2m}$) of
various thousands of kilometers from the $D=0$ point. In these units, the inequality
$ 3400 \leq \Delta L_{2m} \leq 6200 $ is satisfied.

Finally, Fig.~\ref{figu3} corresponds to a third particular direction with two 
$D=0 $ points located at distances $L \simeq 27400 \ km$ and 
$L \simeq 40000 \ km$. This Figure has the same structure as Fig.~\ref{figu2}.
Two peaks and two discontinuities -giving the location of the two $D=0$ points-- 
are observed in the top and middle-bottom panels.
As it follows from the middle-top and bottom panels of Fig.~\ref{figu3},
quantities $\Delta_{R}$ and $|\Delta_{t}|$ take on values of two meters
for 2m-distances ($200 \leq \Delta L_{2m} \leq 500 $) much smaller than those corresponding to the 
$D=0 $ point of Fig.~\ref{figu2}. 
Why does this happen? Further research about this question is in progress.

From all the $D$ values calculated inside the E-sphere, 
it follows that the determinant $D$ does not vanish along any direction
for users with $L < 2.32\times 10^{4} \ km $.
For a given direction, the determinant $D$ 
may vanish once or several times at isolated positions with
$L \geq 2.32 \times 10^{4} \ km $. 
Only along scarce directions, quantity $D$
vanishes more than once. There are also many directions 
without $D=0 $ points. Other 4-tuples and hypersurfaces of constant time,
different from those analyzed here (see Sect.~\ref{sec-4-1}),
lead to very similar conclusions.         

\subsection{S-errors on spherical surfaces concentric with Earth}
\label{sec-4-3} 

\begin{figure*}[tb]
\begin{center}
\resizebox{0.9\textwidth}{!}{%
\includegraphics{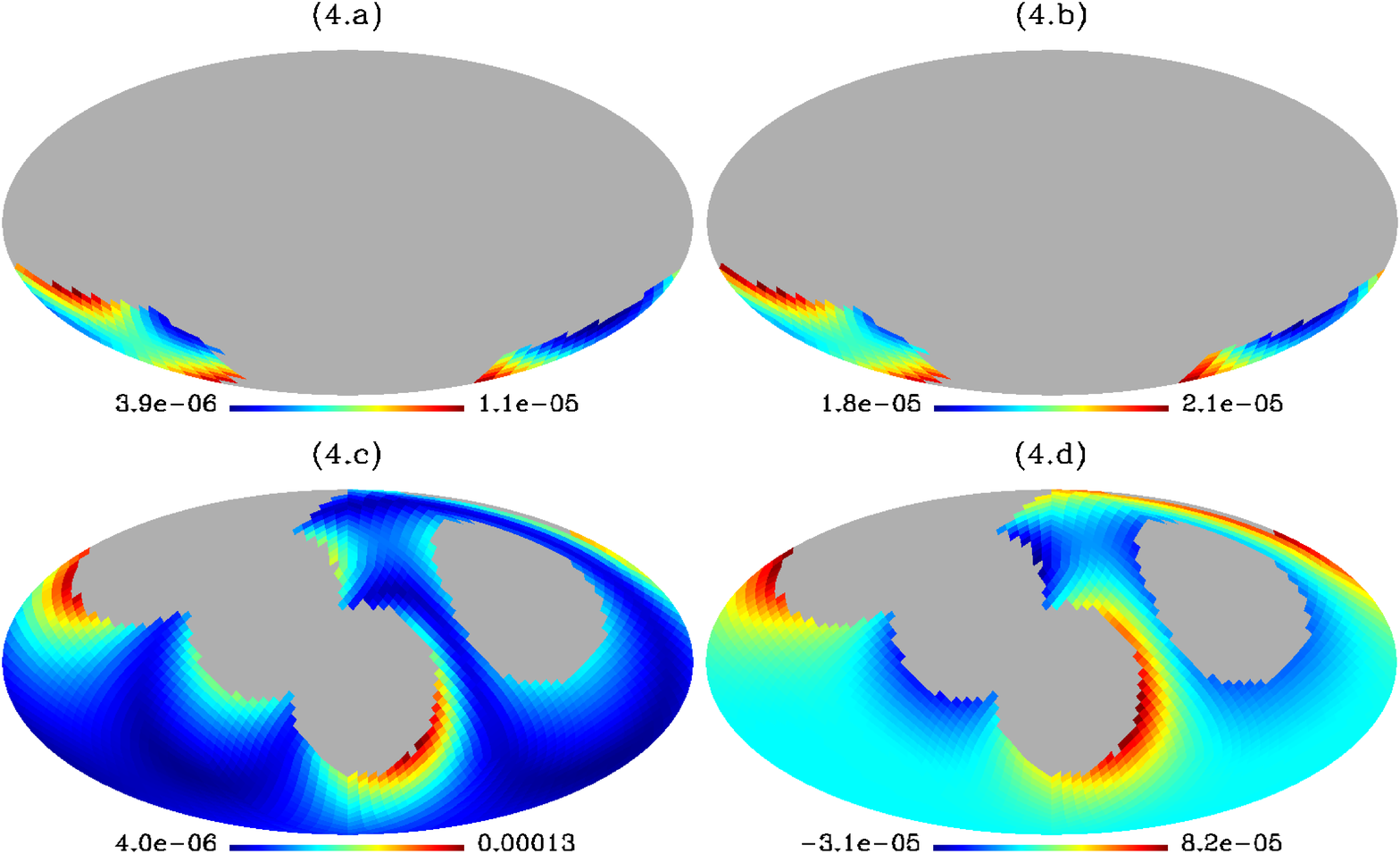}
}
\caption{HEALPIx-mollweide maps of the $\Delta_{R} $ (left) and $\Delta_{t} $ (right)
estimators (in kilometers) on spherical surfaces with different radii. In the top (bottom) panels, the radius 
of the surface --in kilometers-- is $ 6378 = R_{\oplus} $ 
($1.5 \times 10^{4} $). These surfaces are located in the region, around point $E$, 
where the determinant $D$ does not vanish 
}
\label{figu4} 
\end{center}      
\end{figure*}

From top to bottom, Fig.~\ref{figu4} shows the values of $\Delta_{R}$ (left panels)
and $\Delta_{t}$ (right panels) on spherical surfaces concentric with Earth, whose 
radius --in kilometers-- are $ 6378 = R_{\oplus} $ (top) and 
$1.5 \times 10^{4} $ (bottom). 
The two surfaces are inside the E-sphere and located in the region 
where the determinant $D$ does not vanish; hence, 
large values of $\Delta_{R} $ and $\Delta_{t} $ are not expected. 
Actually, all the values displayed in Fig.~\ref{figu4} range from 
$0.39 \ cm$ to $13 \ cm$. In each of the four panels, grey pixels correspond 
to the invisibility points on the concentric spheres.

\subsection{S-errors {\it versus} U-errors}
\label{sec-4-4} 

\begin{figure*}[tb]
\begin{center}
\resizebox{0.9\textwidth}{!}{%
\includegraphics{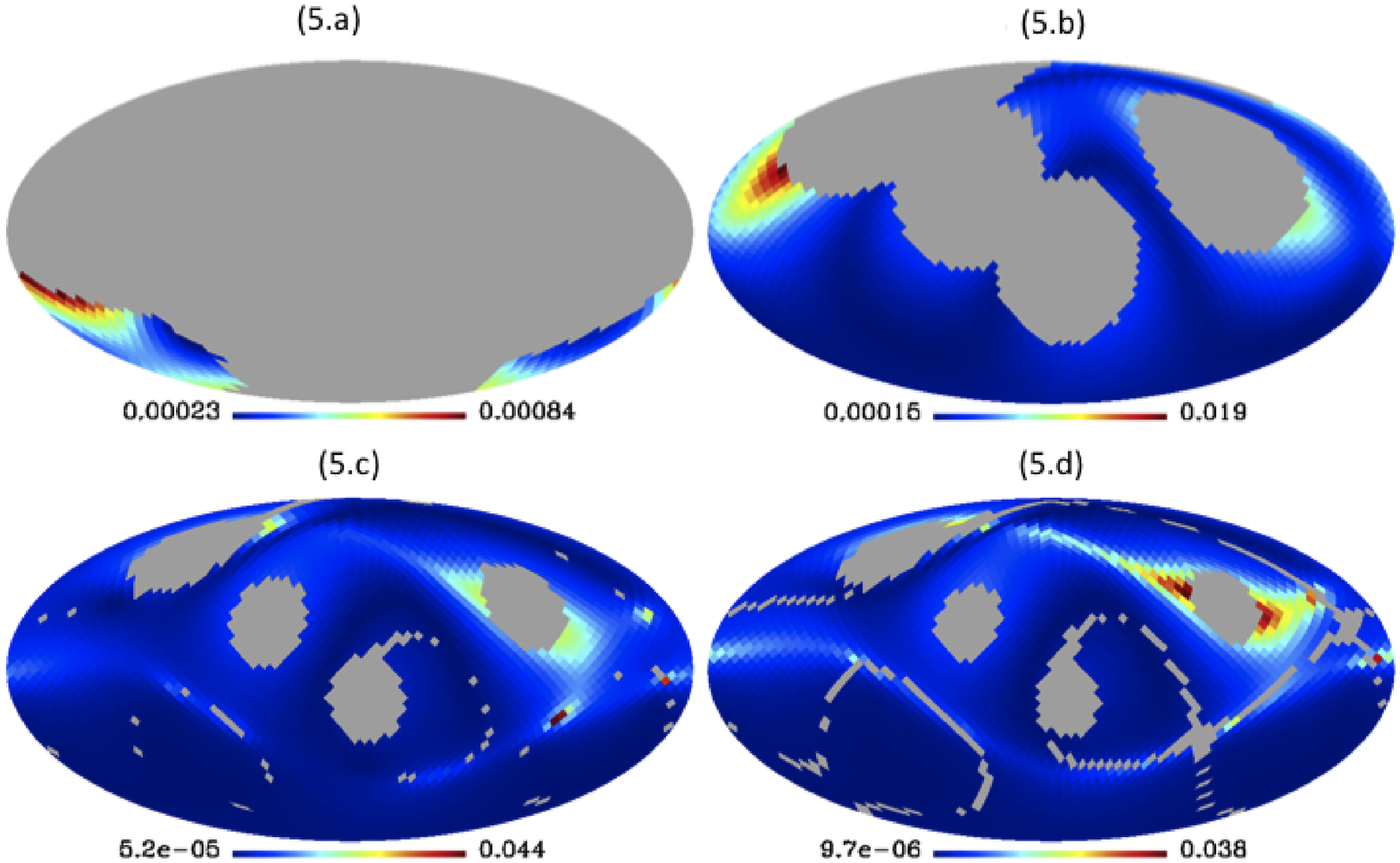}
}
\caption{HEALPIx-mollweide maps. The ratio 
$\xi= \Delta_{RS}/ \Delta_{RU} $, between the 
$\Delta_{R} $ estimators of the S-errors and the U-errors is represented. 
This ratio is 
calculated on four spherical surfaces, concentric with Earth, having radii
[in kilometers] of $ 6378 = R_{\oplus} $ [panel a],
$1.5 \times 10^{4} $ [panel b], $5 \times 10^{4} $ [panel c], and $9 \times 10^{4} $ 
[panel d]
}
\label{figu5} 
\end{center}      
\end{figure*}

The study of this section is based on Fig.~\ref{figu5}.
In the top panels of this Figure, all the grey pixels belong to invisibility 
regions, whereas in the bottom panels, the four big grey spots are the 
invisibility regions and the remaining grey pixels correspond to 
$\Delta_{R} $ values greater than two meters.

Since we have calculated U-errors [see \cite{puc14}] and S-errors at the 
same places inside the E-sphere, the ratio 
$\xi = \Delta_{RS} / \Delta_{RU} $ between the $\Delta_{R} $ estimators
corresponding to the S-errors 
($\Delta_{RS} $) and the U-errors ($\Delta_{RU} $) may be calculated 
at every visibility point. 
In Fig.~\ref{figu5}, the $\xi $ ratios on four 
spheres --concentric with Earth-- are 
represented. In the spheres whose radii are 
$5 \times 10^{4} \ km $ (panel 5.c) and $9 \times 10^{4} \ km $
(panel 5.d), the ratio $\xi $ has only been calculated at 
the points where $\Delta_{RS} $ is smaller than two meters.
So, positions too close to $D=0 $ points are not considered.
In panel 5.a, one easily sees that the $\xi $ values are very 
small on a spherical surface with the Earth's radius, 
where the inequality $ 2.3\times 10^{-4}
< \xi < 8.4 \times 10^{-4} $ is satisfied; hence,
the S-errors are negligible against the U-errors on this surface.
On the spheres with radii of $1.5 \times 10^{4} \ km$ (panel 5.b),
$5 \times 10^{4} \ km $, and $9 \times 10^{4} \ km $, 
the maximum $\xi $ values are $1.9 \times 10^{-2} $, 
$4.4 \times 10^{-2} $, and $3.8 \times 10^{-2} $, respectively;
hence, in these three cases the maximum values of $\xi $
are of the order of $10^{-2} $.
It is also observed (panels 5.b to 5.d) that 
the greatest $\xi$ values (see the color bars) correspond to pixels which are 
close to grey zones. According to the above description of these zones, this means that 
the pixels having the largest $\xi$ values are either close to $D=0$ points or
close to invisibility regions.

\section{Conclusions and discussion} 
\label{sec-5}

In Sect.~\ref{intro}, we have outlined a plan for theoretical developments 
and practical implementations in the field of RPS. We have described the
theoretical foundations [points (i) to (v)], two particular but very 
important RPS approaches [0 and 1 order RPS] and, finally, 
a novel and well structured program to estimate positioning errors
in the framework of RPS [points (1) to (7)]. 
The total positioning error is the addition 
of two contributions: the U-error (satellite world lines) and the 
S-error (photon world lines). 

It is believed that, 
at distances --from Earth-- greater than 
$d_{max} \sim 2 \times 10^{4} \ km$, positioning 
errors are too big and, consequently, 
spacecraft navigation based on GNSS 
is not feasible
(see \cite{den13} and references cited therein). In the context of RPS, 
this is easily understood taking 
into account that all the
$D=0 $ points are located at distances 
greater than $d_{max} $ (see Sect.~\ref{sec-4-2}). 
Since the S-errors diverge (are large) at (close to) the $D=0 $ points, 
it seems that spacecraft navigation is only possible for 
altitudes smaller than $\sim 2 \times 10^{4} \ km$, 
whereas spacecrafts with altitudes greater than $d_{max}$ may approach 
$D=0 $ points where positioning errors are too large. 
Our analysis also suggests the solution to this problem; in fact,  
in a certain 
position, a spacecraft may be close to a $D=0$ point (large positioning errors) for
a certain satellite 4-tuple, but the same position may be far from any point of this type 
for other 4-tuples. This strongly suggests that 
the spacecraft position may be found anywhere inside the E-sphere
by choosing the best 
4-tuple at any moment; thus, the proximity 
to zero points of $D$ might be avoided along the complete world line;
which is necessary to make autonomous spacecraft navigation based on GNSS feasible 
(see also comments in Sect.~\ref{sec-3}).

 In Sect.~\ref{sec-4}, the S-error distribution has been studied 
to conclude that:
(A) in the region surrounding Earth where 
there are no $D=0 $ points,
quantities $\Delta_{R}$ and $|\Delta_{t}|$ take on values smaller than a 
few tens of centimeters, 
(B) the S-errors are large where 
it was expected from the beginning (see the last paragraph of Sect.~\ref{sec-3}),
and (C) positioning quality decays as the 
distance to the Earth's center increases.

Surrounding every $D=0 $ point, there is a region where the S-errors are 
larger than $2 \ m$. The size of these regions ranges from hundreds to thousands of 
kilometers (see Figs.~\ref{figu2} and~\ref{figu3}).

The S-errors have been compared with 
the U-errors for all the users located on four extended spheres.
The U-errors have been estimated as in \cite{puc14}, 
where an amplitude of $10 \ m$ was assumed to 
simulate the deviations between the nominal and real satellite
world lines. Under this assumption, it has been verified that,
for the users with S-errors smaller than two meters,
the ratio $\xi $ defined in Sect.~\ref{sec-4-4} is smaller 
than $\sim 0.05 $ and, moreover, values of the order $10^{-2} $ 
only appear in a few pixels.
This means that the S-errors --below a level of two meters-- are smaller than five 
per cent of the U-errors. In this situation, the 
approach based on the assumption that photons follow 
null geodesics of M-ST may be applied, at least, 
for positioning inside the E-sphere with standard accuracy requirements; however, if the amplitude 
of the satellite deviations becomes --in future-- much smaller than ten meters,
the U-errors (proportional to this amplitude) will be much smaller and,
consequently, 
the S-errors will not be negligible. In such a case, 
it must be assumed that photons move in S-ST. The same occurs if we need 
a high accuracy positioning to deal with some scientific 
problem.

In standard positioning (not RPS), relativistic corrections are usually applied to satellite motions
in GNSS (clock behavior and so on), nevertheless, it is often considered that 
photons move in Minkowski space-time and, consequently, the S-errors produced by the lensing effect 
-due to the Earth's gravitational field- are neglected. Here, a certain method has been used --for the first time-- 
to build up robust maps of these positioning errors. These maps of S-errors will be a practical tool
since they may be used, from now onwards, 
to decide if relativistic corrections --due to lensing-- 
are necessary to study a given problem related with positioning 
on Earth (e.g., carrier signal positioning) or far from Earth (e.g., autonomous spacecraft navigation).  
For the sake of briefness, only a few maps have been presented in Figs.~\ref{figu4} and~\ref{figu5}.

The 1-order RPS may be improved by assuming a slowly rotating
Earth having a realistic mass distribution with small multipoles,
which evolves under the action of astronomical objects such as, 
e.g., the Sun and the Moon, which slightly contribute 
to the local gravitational field 
in the positioning region. Fortunately, there are known metrics to properly 
take into account these improvements (Kerr, Parametrized Post-Newtonian expansions and so on).
A metric of the form $\tilde{g}_{\alpha \beta} = \eta_{\alpha \beta} + s_{\alpha \beta} +
\zeta_{\alpha \beta} $ seems to be suitable to improve on the 
1-order RPS. All the $\zeta_{\alpha \beta}$ quantities 
will be very small
compared with the terms $s_{\alpha \beta}$ (order $GM_{\oplus}/R$).
Since these quantities describe Schwarzschild perturbations, they
are neglected in the 1-order RPS.

Nominal satellite world lines corresponding to the metric
$\tilde{g}_{\alpha \beta}$ may improve the treatment of 
U-errors; however, the term $\zeta_{\alpha \beta} $
is not expected to be significant to describe
photon motions; in fact, since we have proved that
the gravitational influence of the total Earth's mass 
on the photon world lines is small (Schwarzschild against Minkowski),
much smaller gravitational fields --in the positioning zone-- 
such as those due to Earth's multipoles, the Sun, the Moon, and so on,
must produce smaller effects on the photons (negligible in practice). 
The same occurs 
with the effects of Earth's rotation -which is slow- on the photon world lines, which 
are expected to be negligible

A first generalization of the 1-order RPS approach could be obtained by including the solar mass; 
so, the gravitational field would be governed by two sources: a spherically symmetric Earth
and the Sun, which could be considered as a point-like mass. In this situation, a 
system of reference with origin in the Earth-Sun center of mass would be theoretically (physically) 
inertial (quasi-inertial). The gravitation acceleration produced by the Sun around Earth is 
$g_{\odot} \simeq 6 \times 10^{-4} g \simeq 1.4 \times 10^{-2} g_{sat}$, 
where $g$ and $g_{sat}$ are the gravitation accelerations 
produced by Earth on its surface and on the sphere where the Galileo satellites move,
respectively. These numbers show that the solar gravitational field is much smaller
than that of the Earth in the positioning region and, consequently, the influence of 
this field on photon propagation is expected to be negligible. Its effect on 
the motion of the Galileo satellites is being estimated by us, with the approach 
suggested at the top of this paragraph. 

The study of an RPS improving on
the 1-order RPS is beyond the scope of this paper, but our results 
are important to aid the pursuit of this task in the future.
Much research will be necessary before designing and implementing 
good competitive RPS including error estimations.

We can finally estimate the saving in computer time due to 
the use of 0-order RPS calculations (instead of 1-order RPS estimates). 
We have obtained the CPU times taken by our codes to calculate the inertial coordinates $x^{\alpha}$ (outputs)
from the emission coordinates $\tau^{A} $ (inputs). Two codes based on the 0-order and 1-order RPS have been considered. 
They are multiple precision sequential codes and, consequently, calculations are performed with one 
thread. The processor is an Intel(R) Xeon(R) CPU E7-4820 (64 bits) at 2 GHz. By working with 32-40 digits
the resulting times are $\sim 1-2 \ ms$ for the 0-order RPS, and $\sim 20 \ ms$ for the 1-order RPS; hence,
the use of the 0-order RPS --with the exact solution of \cite{col10a}-- leads to a significant saving of computer time; 
nevertheless, 
the 1-order RPS may be also used; in particular, 
for parallel optimized codes running in modern computers. The same would occur for higher order RPS 
based on the same methods, but involving more general metrics, 
nominal satellite world lines, and transfer time functions.
All this strongly suggests that, from the point of view of CPU cost, accurate RPS are as feasible as the 
standard methods used by current GNSS, which are based on relativistic corrections. 
Researches on RPS are motivated by other positioning aspects detailed above such as, e.g., 
systematic error estimates, applications to spacecraft navigation, and so on.

\vspace{1 cm}

ACKNOWLEDGMENTS: This research has been supported by the Spanish
Ministry of {\em Econom\'{\i}a y Competitividad},
MICINN-FEDER project FIS2012-33582.
We thank B. Coll, J.J. Ferrando, and J.A. Morales-Lladosa for
valuable comments.


\begin{thebibliography}{}

\bibitem[Abel and Chaffee(1991)]{abe91} 
Abel, J.S., Chaffee, J.W. Existence and uniqueness of gps solutions. IEEE Trans. Aerosp. Electron. Syst.,
27, 952-956, 1991. 
%
\bibitem[Ashby(2003)]{ash03}
Ashby, N. Relativity in the Global Positioning System. Living Rev. Relativ., 6, (1), 2003. 
%
\bibitem[Bennett et al.(2013)]{ben13}
Bennett, C.L., Larson, D., Weiland, J.L., et al. Nine-year Wilkinson Microwave Anisotropy Probe (WMAP) 
Observations: Final Maps and Results.   
Astrophys. J. Supplement, 208, 1 (54pp), 2013.
%
\bibitem[Bahder(2001)]{bah01} 
Bahder, T.B. Navigation in curved space-time. 
Am. J. Phys. 69, 315-321, 2001.
%
\bibitem[Bini et al.(2008)]{bin08} 
Bini, D., Geralico, A., Ruggiero, M., et al. Emission versus Fermi coordinates: applications to relativistic positioning systems. 
Class. Quantum Grav. 25, 205011 (11pp), 2008.   
%
\bibitem[\v{C}ade\v{z} and Kosti\'c(2005)]{cad05}
\v{C}ade\v{z}, A., Kosti\'c, U. Optics in the Schwarzschild spacetime. Phys. Rev. D, 72, 104024 (10pp), 2005. 
%
\bibitem[\v{C}ade\v{z} et al.(2010)]{cad10} 
\v{C}ade\v{z}, A., Kosti\'c, U., Delva, P.
Mapping the space-time metric
with a global navigation satellite system, final Ariadna report 09/1301,
Advanced Concepts Team, European Space Agency: 1-61, 2010    
%
\bibitem[Chaffee and Abel(1994)]{cha94} 
Chaffee, J.W., Abel, J.S. On the exact solutions of the pseudorange equations. IEEE Trans. Aerosp. Electron. Syst., 
30, 1021-1030, 1994. 
%
\bibitem[Coll et al.(2012)]{col12a} 
Coll, B., Ferrando, J.J., Morales-Lladosa, J.A. Positioning systems in Minkowski space-time: 
Bifurcation problem and observational data. Phys. Rev. D, 
86, 084036 (10pp), 2012.
%
\bibitem[Coll et al.(2010)]{col10a} 
Coll, B., Ferrando, J.J., Morales-Lladosa, J.A. Positioning systems in Minkowski space-time: 
from emission to inertial coordinates. 
Class. Quantum Grav., 27, 065013 (17pp), 2010. 
\bibitem[Coll et al.(2011)]{col11a} 
Coll, B., Ferrando, J.J., Morales-Lladosa, J.A. From emission to inertial coordinates: an analytical approach.  
J. Phys.: Conf. Ser., 314, 0121106 (4pp), 2011.
\bibitem[Delva et al.(2011)]{del11} 
Delva, P., Kosti\'c, U., \v{C}ade\v{z}, A. Numerical modeling of a global navigation satellite system in a 
general relativistic framework. Adv. Space Res., 47, 370-379, 2011.   
\bibitem[Deng et al.(2013)]{den13}
Deng, X.P. et al., Interplanetary spacecraft navigation using pulsars. Adv. Space Res., 52, 1602-1621, 2013.
\bibitem[G\'orski et al.(1999)]{gor99} 
G\'orski, K.M., Hivon, E., Wandelt, B.D. Analysis issues for large CMB data sets.
In Banday, A.J., Sheth R.K. \& Da Costa L. (Eds.),
Proceedings of the MPA/ESO Conference on Evolution of Large Scale Structure,
pp. 37-42, 
Printpartners Ipskamp Enschede. 1999.
arXiv:astro-ph/9812350  
\bibitem[Grafarend and Shan(1996)]{gra96} 
Grafarend, E.W., Shan, J. A closed-form
solution of the nonlinear pseudo-ranging equations (GPS).
Artificial satellites, Planetary geodesy No 28 Special Issue on
the XXX-th Anniversary of the Department of Planetary Geodesy,
Polish Academy of Sciences, Space Research Centre, Warszava,
31, 133-147, 1996.  
%
\bibitem[Langley(1999)]{lan99}
Langley, R.B. Dilution of precision. GPS World, 10, 52-59, 1999.  
\bibitem[Misner et al.(1973)]{mis73} 
Misner, C.W., Thorne, K.S., Wheeler, J.A., Gravitation.
W.H., Freeman and Company, NY. 1973   
\bibitem[Pascual-S\'anchez(2007)]{pas07}
Pascual-S\'anchez, J.F. Introducing relativity in global navigation satellite systems. Ann. Phys., 
16, 258-273, 2007
\bibitem[Press et al.(1999)]{pre99} 
Press, W.H., Teukolski, S.A., Vetterling, W.T., Flannery, B.P. Numerical recipes in fortran 77: the art of
scientific computing.
Cambridge University Press, New York, pp. 355-362, 1999. 
\bibitem[Puchades and S\'aez(2011)]{neu11} 
Puchades, N., S\'aez, D. From emission to inertial coordinates: a numerical approach. J. Phys.: Conf. Ser., 
314, 012107 (4pp), 2011.
\bibitem[Puchades and S\'aez(2012)]{puc12}
Puchades, N., S\'aez, D. Relativistic positioning: four-dimensional numerical approach in Minkowski space-time. 
Astrophys. Space Sci., 341, 631-643, 2012.  
\bibitem[Puchades and S\'aez(2014)]{puc14}
Puchades, N., S\'aez, D. Relativistic positioning: errors due to uncertainties in the satellite world lines. 
Astrophys. Space Sci., 352, 307-320, 2014.  
\bibitem[S\'aez and Puchades(2013)]{sae13}
S\'aez, D., Puchades, N. Relativistic positioning systems: Numerical simulations. 
Acta Futura, 7, 103-110, 2013.
\bibitem[S\'aez and Puchades(2014)]{sae14}
S\'aez, D., Puchades, N. Locating objects away from Earth surface: Positioning accuracy. 
Springer Proceedings in Mathematics \& Statistics, 80, 391-395, 2014.   
\bibitem[San Miguel(2007)]{mig07}
San Miguel, A. Numerical determination of time transfer in general relativity. 
Gen. Relativ. Gravit., 39, 2025-2037, 2007.   
\bibitem[Schmidt(1972)]{sch72} 
Schmidt, R.O. A new approach to geometry of range difference location. IEEE Trans. Aerosp. Electron. Syst., 8, 
821-835, 1972.    
\bibitem[Synge(1931)]{syn31} 
Synge, J.L. A characteristic function in Riemannian space and its application to the solution 
of geodesic triangles. Proc. London Math. Soc., 32, 241-258, 1931.
\bibitem[Teyssandier and Le Poncin-Lafitte(2008)]{tey08} 
Teyssandier, P., Le Poncin-Lafitte, C. General post-Minkowskian expansion of time transfer functions. Class. Quantum Grav., 
25, 145020 (12pp), 2008.

\end{thebibliography}
\end{document}